# Experimental proof of moisture clog through neutron tomography in a porous medium under truly one-directional drying

M. H. Moreira[a,*], S. Dal Pont[b], A. Tengattini[b,c], A. P. Luz[a], V. C. Pandolfelli[a]

[a]Federal University of Sao Carlos, Graduate Program in Materials Science and Engineering (PPGCEM), Rod. Washington Luiz, km 235, 13565-905, São Carlos, SP, Brazil
[b]CNRS, Grenoble INP, 3SR, Universite Grenoble Alpes, 38000 Grenoble, France
[c]Institut Laue Langevin, 71 Avenue des Martyrs, 38000 Grenoble, France

## Abstract

Structural failure of concrete buildings on fire and complete destruction of the monolithic refractory lining during their drying stage are dangerous examples of the effect of explosive spalling on partially saturated porous media. Several observations in both cases indicated the presence of moisture accumulation ahead of the drying front, which are in tune with the most common theories on the explosive spalling of concrete. Previous studies have shown evidence of the existence of this phenomenon, however, they were biased by artifacts and experimental limitations (such as the beam hardening effect and changes in the microstructure of the material due to the presence of pressure and temperature sensors). In the current work, rapid neutron tomography was used to investigate the in-operando drying behavior of a high-alumina refractory castable, proposing a novel experimental layout aimed at a truly one-dimensional drying front. This setup provided more realistic boundary conditions, such as the behavior of a larger wall heated from one of its sides, while also preventing some nonphysical artifacts (notably beam hardening). By eliminating these aspects, a direct proof that moisture accumulates ahead of the drying front was obtained. This work also lays the basis for further studies focusing on the response sensitivity analysis to boundary conditions and other parameters (e.g., heating rates and properties of the sample related to the moisture clog formation), as well as useful data for the validation and characterization stages of numerical models of partially saturated porous media.

**Keywords:** Concrete Spalling, Refractory Castable, Drying Front, Neutron Tomography, High Temperature

## 1. Introduction

The behavior of saturated porous ceramic media during heating is of great importance for several distinct applications, ranging from studying the fire safety of Portland cement concrete structures [1, 2, 3, 4] to the drying of refractory castables [5, 6, 7, 8]. Due to the high heating rates, low permeability of the medium and high thermomechanical stresses, the explosive spalling phenomenon might take place, leading to cracks, damage and even complete failure of such structures [3, 9, 7, 5, 10].

Because of the size scale of such structures, using numerical modelling is of great interest, as full-scale measurements are often limited and expensive [11, 12, 13, 14]. Nevertheless, to properly apply these numerical tools for studying and designing purposes, validating them by comparing the predictions with experimentally measured values, is required. The most common test used for this purpose, the PTM (pressure, temperature, and mass evolution), is based on using thermocouples and pressure sensors placed inside prismatic samples, which are heated from one of its sides [15, 6, 16].

Dauti et al. relied on neutron tomography to evaluate whether and how these sensors affect the microstructure of high performance Portland cement concrete [17]. It was observed that placing a thermocouple with only 0.25 mm in diameter inside a concrete sample was capable of inducing air bubbles formation, a problem that could be greater in the case of thicker pressure sensors, as the ones commonly used for the PTM tests [17]. As a consequence, care must be taken when considering inserting these devices in the structure of the analyzed materials, as they might directly affect the measurements.

This drawback favors using non-destructive imaging techniques such as X-ray tomography, nuclear magnetic resonance or neutron tomography itself as a way to directly assess the water distribution inside a porous media [18, 4, 17, 19]. Specifically, neutron imaging has the advantage of providing an ideal contrast between wet and dry regions, as neutrons are highly sensitive to hydrogen atoms (much more abundant in the wetter portions of the sample, making them more neutron opaque) [4]. It also enables the 3D reconstruction of the water content distribution, a feature that is not available for regular NMR based techniques proposed in the literature [18, 20].

Using such a method, Dauti et al. was able to observe the effect of the aggregate size on the moisture mass transfer inside the samples, as well as to validate its mesoscopic thermohygromechanical model based on the collected data [4, 13]. In their work, an increase of neutron attenuation ahead of the drying front was also detected, which suggested the occurrence of the moisture clog phenomenon, a key aspect to the current understanding of explosive spalling. The presence of this moisture clog is thought to follow the condensation of liquid water at the cold surface of samples being heated from a single direction [3, 7].

An increase of neutron absorption in the innermost area of the sample was also detected, which is a further indication of the moisture clog phenomenon, which may directly affect the likelihood of facing explosive spalling events. This theory is supported by physical observations, such as the condensation of liquid water at the cold surface of samples that is heated from a single direction [3, 7].

Although common during the heating of large volumes of porous media, the moisture clog was not directly observed nor quantitatively described in laboratory-based studies, indicating the possibility of a size-dependency of the process.

One of the greatest challenges for using neutron tomography for studying this process is perhaps the limited ideal sample size that can be tested which is in the centimeter range [4] due to the transmission (because of the high hydrogen content in the concrete or refractories) and also the resolution required to capture the key processes. This implies that for the tests to be truly representative, the boundary conditions should be close to the ideal, and, in principle, expressing those attained for larger structures.

In this context, one-dimensional drying can be argued to be a reasonable representation of the homogeneous drying of a larger wall. Additionally, a truly 1-D front can help to withdraw doubts arising from some artifacts present in the tomographic images, most notably beam hardening [4, 17, 21] which can result in a slight spurious change of attenuation in the radial direction. While generally small, radial drying can induce a spurious increase in attenuation that could be misinterpreted as moisture accumulation, as analyzed in detail in [4]. One-dimensional boundary conditions can withdraw the influence that this artifact may have on a rigorous analysis of the moisture mass transfer and its accumulation ahead of the drying front. This is of particular importance when using these results for the validation of numerical models (without the drawbacks of the PTM test) and to obtain properties of the materials, such as the retention curves (also known as the sorption isotherm) and the intrinsic permeability [4].

Trying to provide a unidimensional drying front, Tengattini et al. analyzed the effect of using distinct casings based on titanium and quartz, which yielded results less prone to the beam hardening effect [21]. Nevertheless, the so-called boundary effects were still present in such cases and lateral drying was also observed.

The current work proposes a materials' selection-based method for the optimization of casings to ensure a one-dimensional drying front during neutron tomography tests and, consequently, attest and quantify the moisture accumulation previously only mentioned by other studies [4, 13, 21], which were subject to the beam hardening effect. As the mass transfer is reduced to a single direction phenomenon, the selected setup can be used to effectively describe the behavior of a solid wall of refractory castables, a class of materials not previously studied with full-field techniques. These results can validate the presence of a moisture clog even in small-scale samples, providing a strong indication that this phenomenon is directly linked to the explosive spalling events.

## 2. Material and Methods

### 2.1. Evaluated castable composition and experimental setup

A self-flowing high-alumina refractory castable containing 5 wt% of calcium aluminate cement as a binder (thus, named here as 5CAC) was designed according to Andreasen's particle packing model and considering a distribution coefficient q = 0.21. Table 1 shows more details of the evaluated composition.

The refractory composition was homogenized for 1 min and mixed with water for an additional 3 min in a paddle mixer capable of measuring the torque applied [22]. Afterwards, cylindrical samples (50mm in height and 33mm in diameter) were cast in PVC molds (samples without casing) or in a mullite ceramic casing (samples with casing). Then, they were cured at 30C with 80% of relative humidity in a climatic chamber for 24h.

Table 1: High-alumina refractory castable composition.

| Raw materials | | Compositions (wt.%) | Compositions (vol.%) |
|---|---|---|---|
| Tabular alumina | AT 6-3 | 18 | 16.2 |
| | AT 3-1 | 10 | 9.0 |
| | AT 1-0.5 | 11 | 9.9 |
| | AT 0.6-0.2 | 9 | 8.1 |
| | AT 0.2-0 | 16 | 14.3 |
| | AT < 45 | 10 | 9.0 |
| Calcined and reactive alumina | CL370 | 11 | 11.5 |
| | CT3000SG | 10 | 16.0 |
| Calcium Aluminate Cement | Secar 71 | 5 | 6 |
| Water | Distilled water | 4.5 | 16.1 |
| Dispersant | Castament FS60 | 0.2 | 0.01 |

This composition was selected to be representative of refractory castables, which have not, to the best of the authors' knowledge, been previously studied in terms of their full- eld moisture migration, and additionally, to highlight how the choice of the impermeable casing needs to take into account the properties of the samples.

An ideal casing material needs to be transparent to neutrons, to withstand the high temperatures imposed during drying and to be impermeable. There are multiple candidates that partially satisfy such requirements. Nonetheless, one major point that also needs to be considered is the thermal expansion of the material. For instance, casings based on materials with higher thermal expansion coefficients will expand more than the evaluated samples, leading to vapor release in the radial direction. On the other hand, when the thermal expansion of the casing is much lower, the induced mechanical stresses might result in cracks, which would spoil its main purpose.

Figure 1 presents the linear thermal expansion of different materials and the designed refractory castable in the range of 100C to 550C (typical values attained for the neutron tomography experiments [4]).

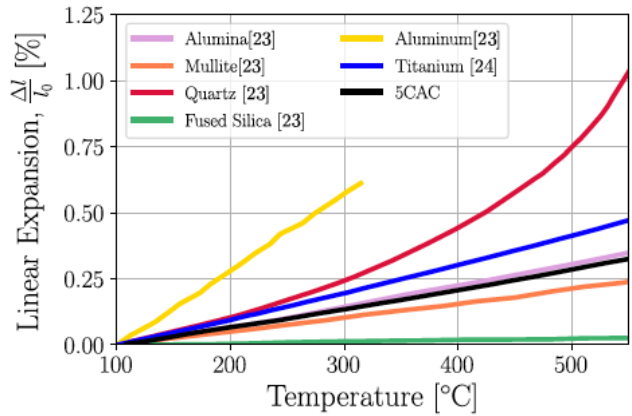

Figure 1: Linear thermal expansion of different material candidates for casings of neutron tomography tests. Adapted from [23, 24].

It can be observed that previously tested materials such as titanium and quartz present higher linear expansions than those obtained for the 5CAC composition. Aluminum is also not a suitable material as it shows the highest thermal expansion coefficient. Fused silica has almost no thermal expansion in this temperature range, making it prone to cracks during heating, from where moisture could be released. Sintered alumina and mullite are then in this case the most suitable options, and were analyzed in this work. As both casings performed similarly, the results presented are based on the mullite sintered ceramic. The geometry of the casing was an open-ended cylinder with the same height of the samples (with 50mm in height and 33mm in diameter).

The setup for the neutron tomography tests in the NeXT equipment follows the same approach already described by Dauti et al. and Tengattini et al. [4, 17, 21], where the samples were wrapped up in rock wool to thermally insulate the sides and the bottom of the sample. The biggest difference consists of the fact that the samples were directly cast inside the ceramic casings, as depicted in Figure 2.

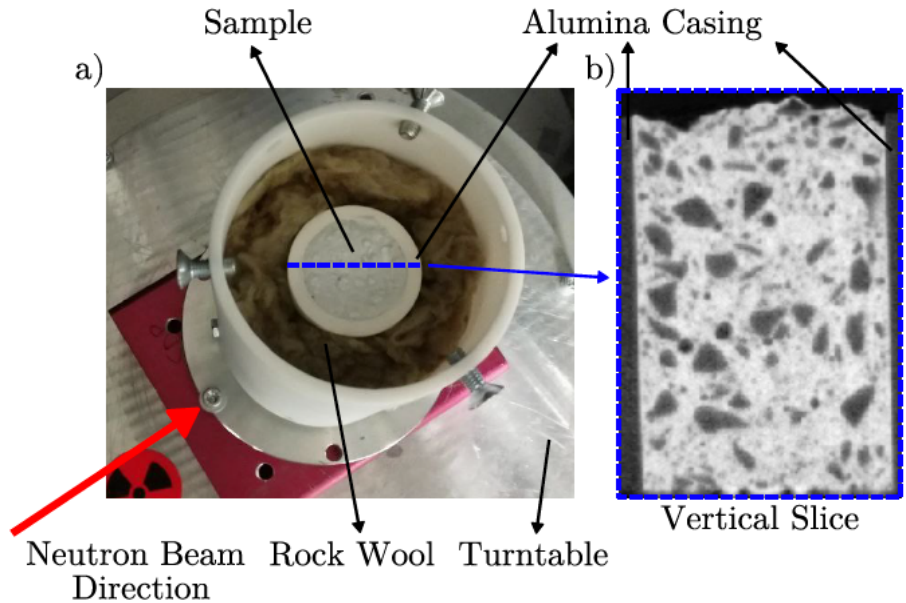

Figure 2: (a) Experimental set up for the neutron tomography test in the NeXT equipment at Institut Laue-Langevin (ILL) and (b) vertical slice of the 3D reconstruction of the neutron tomography of the sample showing the alumina casings.

The Neutron tomographies were acquired using the NeXT instrument at the Institut Laue-Lavengin (ILL) [25]. The true spatial resolution was 160 µm whereas the time interval between two successive tomographies was 58.5s, as described in the tests reported by Tengattini et al. [21]. The setup for the heating tests follows the same approach already described by Dauti et al. and Tengattini et al. [4, 17, 21], with a heating rate of the infrared heater of 10C/min up to 500C, reaching a plateau afterwards.

### 2.2. Image analysis

Firstly, the 3D volumes were reconstructed using the Feldkamp filtered back projection using a commercial reconstruction software (X-act, by RX Solutions). The volumes were described by 300 slices in the cylinder's depth direction (z), and saved as TIFF images with 200 x 200 pixels. Thus, each tomography consisted of an array of dimensions (200, 200, 300) and an additional one related to the time step, resulting in a data array for each experimente with dimensions (t, 200, 200, 300).

Because the main focus of the work was to analyze the behavior of the castable matrix (cement + calcined and reactive aluminas) rather than the aggregates, the latter were removed from the images via a process called\segmentation". In this case, the aggregates were simply detected due to their particularly low neutron attenuation (alumina has a very low neutron opacity), so that any voxels (3D pixels) of the 3D volume displaying an attenuation value below a given threshold were masked out from all images and consequently the analyses. This segmentation was carried out at the reference state of the sample (before the drying process), and more specifically on an image resulting from the average of the first ten tomographies (where virtually no drying occurred), as in [4]. It should be noted that, given the 160 µm resolution of the images, only the larger aggregates could be removed by this method. The main purpose of this aggregate removal was to avoid affecting the drying ratio calculation (presented next) due to the inhomogeneous distribution of the larger aggregates. It is assumed here that the aggregates below the resolution range are sufficiently homogeneously distributed to not affect the proposed measurements.

It should be noted that when considering the particle size distribution of all components of the 5CAC composition, only a fraction of the aggregates (around 25.2 vol.%) were indeed detectable at the spatial resolution range used for the neutron tomographies (that starts with 160 µm), as seen in Figure 3. The only aggregates that completely fall inside the detection range are the tabular alumina AT 6-3 and AT 3-1.

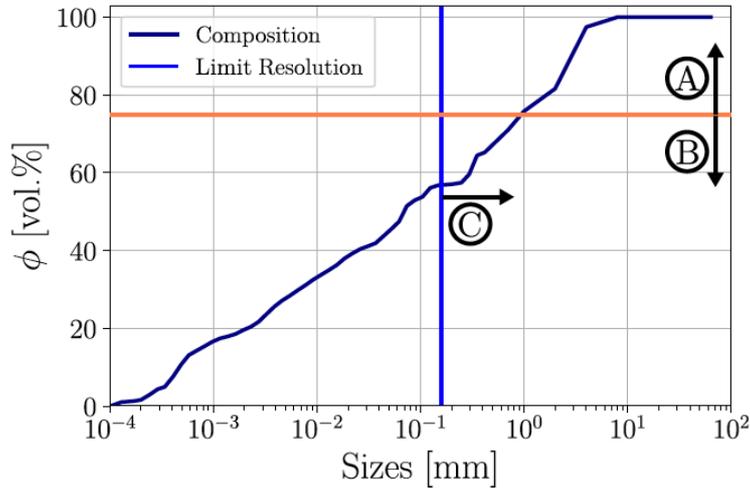

Figure 3: 5CAC composition particle size distribution. The aggregates above the resolution could be removed by the segmentation process, whereas the ones below (< 0:160mm) were assumed to be sufficiently homogeneous to be part of the matrix. Arrow \A" describes the volume fraction that lies completely in the detection range (AT 6-3 and AT 3-1 fractions corresponding to 25.2 vol.%), whereas \B" is the remaining part of the composition and \C" describes the detectable size range.

### 2.2.1. Relative changes in water distribution

As a direct manner to compare the drying front profiles, the initial results can be set as a reference condition, in which the water content is assumed to be evenly distributed throughout the sample. This reference state was obtained by the median of the first ten tomographies.

To study the drying front, the difference can be analysed between the neutron attenuation maps at a particular time t, and the reference state where the moisture distribution is assumed to be homogeneous, given that the only change in neutron attenuation is caused by the moisture migration. The relative difference, $\psi(x; t)$, at any given moment, t, and a spatial position x = (x; y; z), can be calculated considering Equation 1:

$$\psi(\mathbf{x}, t) = \frac{\left(I(\mathbf{x}, t) - \hat{I}(\mathbf{x}, t_0)\right)}{\hat{I}(\mathbf{x}, t_0)} \times 100\% \tag{1}$$

where $\hat{I}(x; t_0)$ is the aforementioned reference state.

### 2.2.2. Drying ratio evolution

The second quantity used for describing the drying process in the present work was the drying ratio evolution. This quantity was obtained by considering the ratio between the number of voxels that are dry at a given moment (with intensities lower than $I_{wet}$) and the initial number of matrix fraction voxels (fine components of the composition which are not aggregates, with intensities higher than $I_{aggs}$), as shown in Equation 2.

$$\frac{V_{dry}}{V_{matrix}} = \frac{\{\sum_i i \mid I(\mathbf{x_i}, t) < I_{wet}\}}{\{\sum_i i \mid \hat{I}(\mathbf{x_i}, t_0) > I_{wet}\}} \tag{2}$$

The attained result will depend on the chosen threshold. A sensitivity analysis confirms that, while this choice affects the exact volume fractions, it does not (within a reasonable range) affect their trends, as it was already the case in previous studies [21, 4]. Regarding the latter issue, aggregates that were not removed by the mask developed might be accounted for as an initial amount of matrix paste, when in fact, they belong to the finer fractions of the aggregates used in the refractory castable composition (see Table 1). As the composition is the same for both samples conditions (with or without casing), the relative comparison is still valid. To quantify the relative performance in terms of its capacity to impose a one-dimensional mass transfer, the drying ratio was calculated considering both the whole sample and accounting only the core of the sample (1/3 of the sample's diameter).

### 2.2.3. Drying front position

To validate the previous approach and obtain the drying front position regardless of the arbitrary thresholds, a second procedure was proposed to directly calculate the drying front position based on its gradient, as presented in Figure 4. The basic idea was to obtain the front position by finding the minimum of the gradient of the intensity on the axial direction. Thus, firstly, for each tomograph, the arbitrary intensity values were averaged in the radial direction (on each radial slice). This procedure reduced the result to a single array with dimension (300, 1). The result can be seen in the center plot in Figure 4.

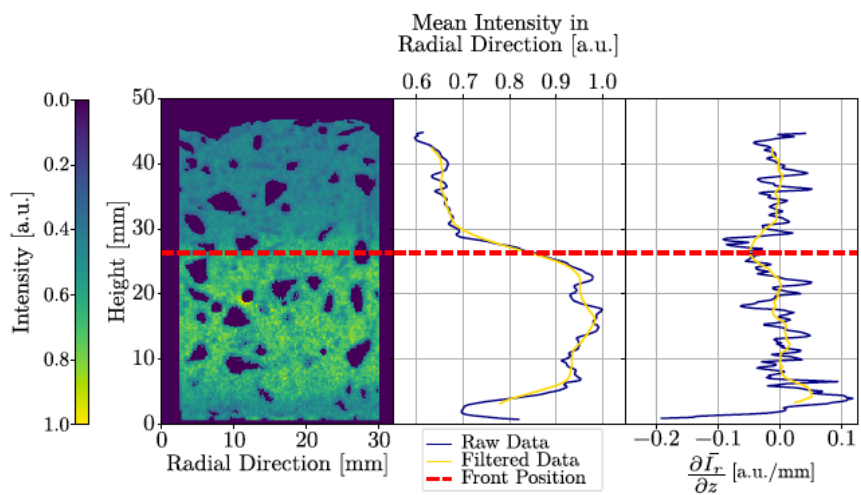

Figure 4: Drying front position detection algorithm. Firstly, the mean of the intensity value on the radial direction was calculated, then a Savitzky-Golay filter was used to reduce the noise of the data. Finally, the minimum of the derivative of the radial mean intensity with respect to the height defined the drying front position, which is represented by the red dashed line (notice how it is defined by the minimum of the yellow line on the derivative plot).

Due to the presence of aggregates and to the nature of the data, some noise could be observed, which can affect the gradient based algorithm. Thus, a digital filter, based on the Savitzky-Golay one [26], was applied to the data before the numerical differentiation. This procedure provided a smoother result, as shown by the plot on the right side of Figure 4. To compare both results obtained by the drying ratio and drying front position, the drying percentage was then calculated as the fraction of the sample that fulfills the dry condition.

## 3. Results and Discussion

The effects of the casing on the drying behavior of 5CAC refractory are clear when based on the relative changes in water distribution over time, reported in Figures 5 and 6. Figure 5 highlights how the casing-free sample was almost completely dry after 63 minutes (see regions in dark blue), whereas the one with the ceramic shell was only partially dried after the same heating time. (Figure 6). This can be explained by the lateral drying, which started around the 31-minute frame for the casing-free sample. This mass flux in the radial direction led to a curvature of the drying front (in the 3D volume, this is actually a dome-shaped front) that was not observed for the sample with the ceramic casing (at boundary).

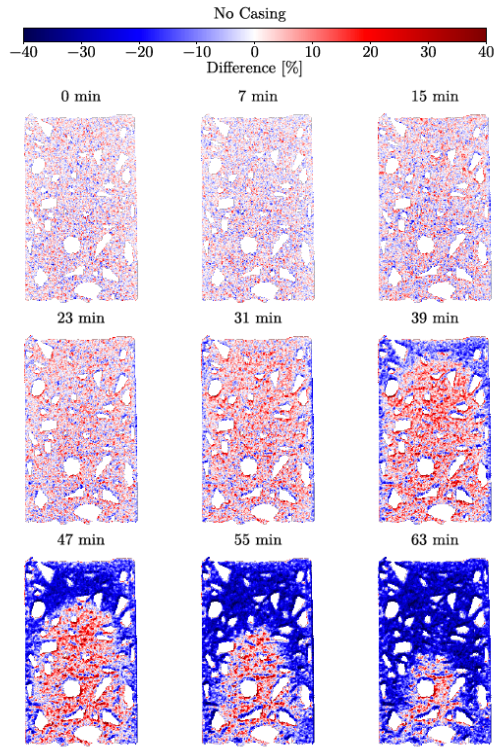

Figure 5: Relative changes in water distribution over time for sample without casing.

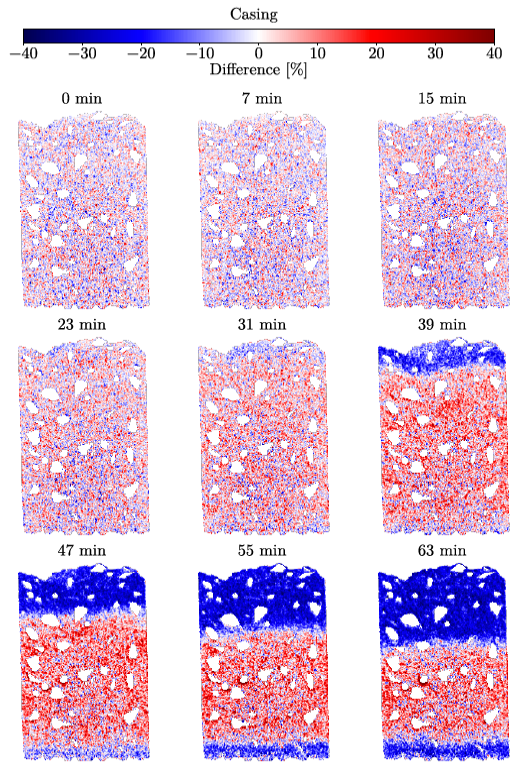

Figure 6: Relative changes in water distribution with time for sample with the mullite casing.

Regarding the moisture accumulation ahead of the drying front (regions in red in Figures 5 and 6), the small quantity of water observed at the core of the sample without the casing could be argued to be either physical (i.e. moisture accumulation) or related to the beam hardening effect, resulting in an artificial increase in the detected intensity in such regions [4, 17, 21]. However, the sample with the casing presented a  at drying front, suggesting that the moisture accumulation (more pronounced than in the casing-free sample) is actually physical.

Upon closer inspection, two drying fronts inside the sample with the casing were identified, a faster one moving from the top surface (hot region) of the cylinder towards its middle and a second one shifting from the cold bottom surface to the middle of the sample. This behavior might be related to the fact that, as water cannot escape in the radial direction, it will start to be released via the cold surface of the sample. In an explosive spalling event, the moisture clog would limit this flux due to the water condensation at the colder portions of the material, saturating the pores, leading to the pressurization and, consequently, damaging the structure [9, 3].

This hypothesis agrees with the observed behavior for large-scale ceramic lining walls when they are subjected to fire tests, where liquid water is seen pouring out of the cold surface [7, 3, 27] and also using neutron tomography for the sample with the ceramic casing.

It should be noted, however, that for the current experimental setup, although the casing is wrapped up by a fiber insulating ceramic, heat can move around the sample, therefore this secondary drying front can be related to some heating of the bottom surface. Furthermore, studies focusing on the thermal profile as obtained through thermocouples will be needed to confirm or dismiss this hypothesis as the explanation of the observed secondary drying front.

Therefore, Figure 7 presents the dry castable matrix fraction of each sample as a function of time, described in Section 2.2.2. As observed, the drying ratio approach agrees with the drying front position results for both evaluated conditions (samples with and without the ceramic casing). Minor fluctuations on the absolute values could be ascribed to the selected threshold values, which were required for the calculation of the drying ratio (Equation 2).

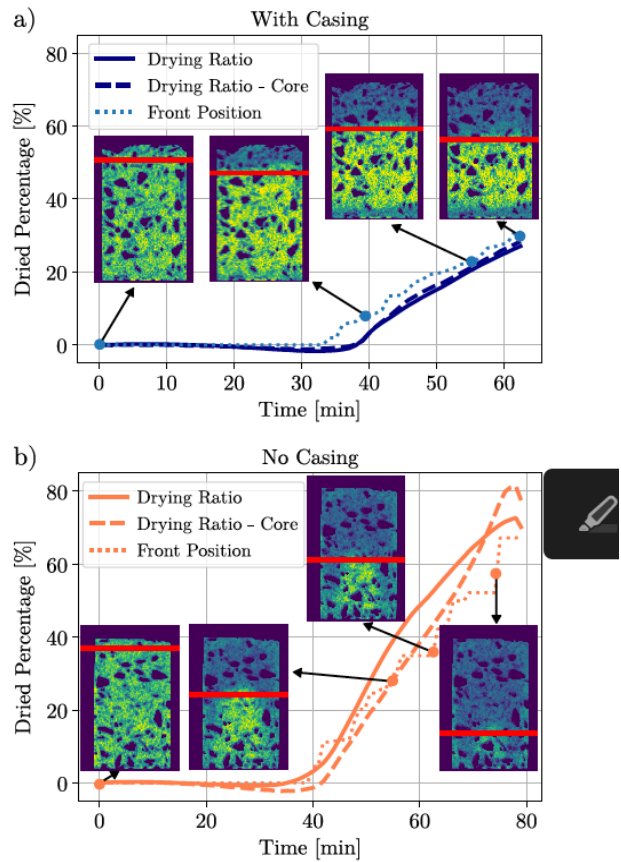

Figure 7: Quantification of the drying evolution of 5CAC castable for samples with casing (a) and without the casing (b). Such results were obtained using the two proposed methods, i) the drying ratio calculation (based on the ratio between dried voxels and total matrix voxels, as described in Section 2.2.2) for the whole sample or its core and ii) the front position (using a gradient based algorithm on the pro le of average moisture content along the height of the samples,

Section 2.2.3, normalized by the height of the sample). Insets show the axial slices of the tomography at different moments, with the drying front position highlighted by the red line, for different times of the drying process.

The insets show the drying front detected at specific times indicated by the dot markers on the plots. Once again, it was identified at drying front was identified when using the proper ceramic casing. This observation can be confirmed by comparing the drying ratio computations for the whole sample (continuous lines) and only for its core region (dashed results) in Figure 7 (b). The observed differences between the core and the entire sample with the casing are negligible (smaller than 1.5%) - indicating the effectiveness of the casing in preventing lateral drying, whereas the caseless sample results differ by up to 10%, highlighting considerable radial mass transport.

To carry out a more detailed analysis on the effect of the ceramic casing on the drying behavior of the castable's samples, a 2D contour map can be drawn to describe, in the radial direction, the average of the difference of water content between the reference state (the median of the first ten tomographies) and at any moment during the test (in the X-axis) as a function of the samples' height (depth, in the Y-axis), as described in Figure 8.

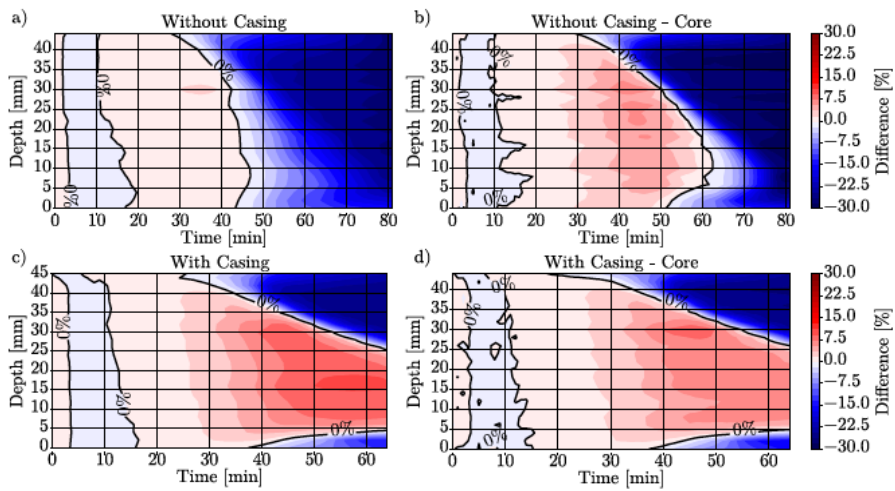

Figure 8: Average in the radial direction of the water difference distribution as a function of time for the sample without the ceramic casing, (a) and (b), and for the one with casing (c) and (d). The radial mean value was calculated for the whole sample's diameter in (a) and (c) and only for the core section in (b) and (d). Higher values of difference (dark red) indicate water accumulation, whereas lower values (dark blue) represent drying. The results of the first ten tomographs (roughly the first 10 minutes of the test) are not described here, as they are the reference state.

The initial decrease of relative difference (which indicates a decrease in the water content) observed for both samples from the middle to the bottom part of the samples, with and without the casing, between 10 and 20 minutes, could not be physically explained and could be related to thermal expansion effects.

Water accumulation (regions in darker red) was also detectable for the case-free sample when only the core (Figure 8 (b)) was considered for the time range of 40 and 50 minutes and between 20mm and 35mm depths. In this case, the drying at the bottom of the sample (colder side) only started after 50 minutes.

For the sample with the ceramic casing, the same pattern was observed when analyzing the whole sample or only its core (Figure 8 (c) and (d)), that is, two distinct drying fronts, separated by a region of moisture accumulation. This moisture clog increases its intensity with time, from 50 minutes and onwards, indicating a higher water accumulation for depths in the range of 5mm and 25mm.

To further investigate the behavior of different regions of the sample, the evolution of the normalized average gray value (i.e. the normalized mean intensity of each voxel) at specific positions can be studied, as shown in Figure 9. Focusing, for example, on the sample drying status after 60 minutes, it can be observed how the moisture accumulation is much more pronounced for the sample with the casing and how the sample without it dries faster.

Moreover, due to the absence of mass transport in the radial direction, the drying evolution at the bottom surface of the sample with the ceramic casing is more intense and faster, as shown in Figure 6. These results indicate how the drying phenomenon is sensitive to the boundary conditions and how different configurations impact the release of water from a porous medium. These observations explain the difficulties of predicting the drying behavior of refractory castables and for the Portland cement concrete under fire, as complex geometries and boundary conditions may directly impact the mass transfer of water.

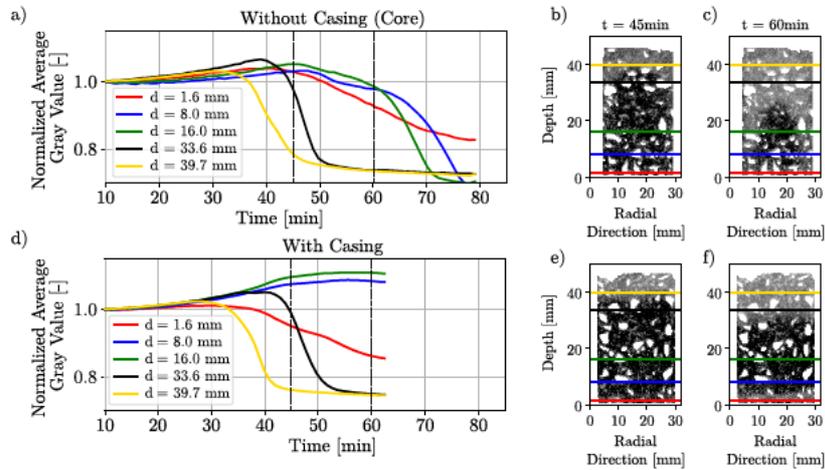

Figure 9: Normalized average gray value evolution at specific depths of the high-alumina castable's samples (a) without the casing and (d) with the casing. The slices of the tomographs at 45 and 60 minutes along with the different depths considered in the plots are represented in (b), (c), (e), and (f). The results of the first ten tomographs (roughly the first 10 minutes of the test) are not described here, as they are the reference state.

This, in turn, directly affects the pressurization of vapor, and consequently the stress state of the structure, altering how the explosive spalling occurs. The technique provided herein can, thus, be used to compare and develop novel strategies for reducing the risks of facing such damaging events, regardless of the application: either it is the drying of refractory castables or fire accidents.

Finally, the approach presented imposes boundary conditions that yield an one-directional mass transport inside the sample, which is representative of a section of a flat wall with the same thickness as the samples' heights, as schematized in Figure 10. It follows that the results and findings can be considered and extrapolated to bigger structures and geometries, providing a solution to one of the greatest drawbacks for using Neutron Tomography to analyze the drying behavior of concrete, which was the size limitation of the tested samples.

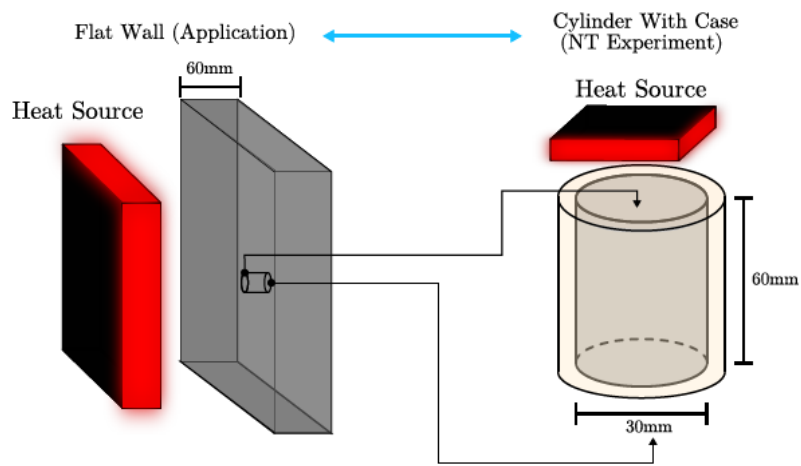

Figure 10: Geometry and boundary conditions equivalence between a at wall commonly found in the cases of interest when studying the drying and fire safety, and the cylindrical neutron tomography sample due to the one dimensional mass transport provided by the casing.

## 4. Conclusions

This work proposed a method for materials selection to be used to impose a truly unidimensional drying under rapid neutron tomography evaluation. In order to attain this requirement, the ceramic casing needs to present neutron transparency, low permeability and mostly important a thermal expansion coefficient similar to the sample's one. Considering the adopted refractory castable composition, mullite and alumina casings were suitable options.

Comparing the drying behavior of the samples tested with and without the mullite casing pointed out that the drying front was clearly distinct. For the sample without the ceramic casing, an intense mass transport in the radial direction and thus a curved front, could be detected, whereas the casing-containing one presented a flat one-dimensional boundary. Besides the intrinsic interest of this 1D mass transport, the likelihood of artifacts, such as the beam hardening effect, was eliminated, and thus the present work attested the presence of water accumulation ahead of the drying front, indicating that even for smaller samples, such a phenomenon can be observed. This also reinforces the moisture clog hypothesis which explains the explosive spalling of Portland cement concrete under fire and the accidents during the drying of refractory castables.

Another important aspect was the drying that took place at the bottom surface of the sample with the casing, which is of great importance for the heat-up process of refractory linings, where an appropriate number and size of drain holes (commonly known as weep holes) need to be designed to maximize the drying step, because if water cannot be released through the cooler surface, the resulting scenario is more likely to yield an explosive spalling event, besides decreasing the efficiency of the drying.

The current methodology lays an important basis for future experiments both for Portland cement concrete and refractory castables, enabling a reliable quantitative analysis that could be used to deduce properties such as the sorption isotherm, the intrinsic permeability of the material and also to validate numerical models. Another important application is the development and benchmark of novel additives and solutions to minimize explosive spalling. All these rely on boundary conditions capable of realistically representing engineering problems such as flat lining walls undergoing heating, providing extrapolations of such findings for actual industrial and fire accident scenarios.

## 5. Acknowledgments

This study was financed in part by the Coordenação de Aperfeiçoamento de Pessoal de Nível Superior - Brasil (CAPES) - Finance Code 001. The authors would like to thank the Fundação de Amparo a Pesquisa do Estado de São Paulo - FAPESP (grant number: 2021/00251-0 and 2019/07996-0) and the Conselho Nacional de Desenvolvimento Científico e Tecnologico - CNPq (grant number: 303324/2019-8). Finally, the authors are greatly thankful for FIRE support to carry out this work. The raw datasets collected at NeXT-Grenoble are available at: https://doi.ill.fr/10.5291/ILL-DATA.UGA-112

## References

[1] Bazant ZP, Thonguthai W. Pore pressure and drying of concrete at high temperature. ASCE J Eng Mech Div. 1978;104(5):1059{1079.

[2] Gawin D, Majorana C, Schreer B. Numerical analysis of hygro-thermal behaviour and damage of concrete at high temperature. Mechanics of Cohesive-frictional Materials: An International Journal on Experiments, Modelling and Computation of Materials and Structures. 1999;4(1):37{74.

[3] Jansson R. Fire spalling of concrete: theoretical and experimental studies [Ph.D. thesis]. KTH Royal Institute of Technology; 2013.

[4] Dauti D, Tengattini A, Dal Pont S, Toropovs N, Brifaut M, Weber B. Analysis of moisture migration in concrete at high temperature through in-situ neutron tomography. Cement and Concrete Research. 2018;111:41{55.

[5] Gong ZX, Mujumdar AS. The influence of an impermeable surface on pore steam pressure during drying of refractory concrete slabs. International Journal of Heat and Mass Transfer. 1995;38(7):1297{1303.

[6] Fey KG, Riehl I, Wulf R, Gross U. Experimental and numerical investigation of the first heat-up of refractory concrete. International Journal of Thermal Sciences. 2016;100:108{125. Available from: http://dx.doi.org/10.1016/j.ijthermalsci.2015.09.010.

[7] Palmer G, Cobos J, Howes T. The Accelerated Drying of Refractory Concrete - Part 1 : A Review of Current Understanding. Refractories Worldforum. 2014;6(2):75{83.

[8] Luz A, Moreira M, Braulio M, Parr C, Pandolfelli V. Drying behavior of dense refractory ceramic castables. Part 1 { General aspects and experimental techniques used to assess water removal. Ceramics International. 2021;47(16):22246{22268.

[9] Ozawa M, Uchida S, Kamada T, Morimoto H. Study of mechanisms of explosive spalling in high-strength concrete at high temperatures using acoustic emission. Construction and Building Materials. 2012;37:621{628.

[10] Innocentini MD, Cardoso FA, Akyioshi MM, Pandolfelli VC. Drying Stages during the Heating of High-Alumina, Ultra-Low-Cement Refractory Castables. Journal of the American Ceramic Society. 2003;86(7):1146{1148.

[11] Gawin D, Pesavento F, Schrefler BA. What physical phenomena can be neglected when modelling concrete at high temperature? A comparative study. Part 2: Comparison between models. International Journal of Solids and Structures. 2011;48(13):1945{1961. Available from: http://dx.doi.org/10.1016/j.ijsolstr.2011.03.003.

[12] Bazant ZP, Jirasek M. Creep and hygrothermal effects in concrete structures. vol. 225. The Netherlands: Springer: Dordrecht; 2018.

[13] Dauti D, Dal Pont S, Bri aut M,Weber B. Modeling of 3D moisture distribution in heated concrete: From continuum towards mesoscopic approach. International Journal of Heat and Mass Transfer. 2019;134:1137{1152.

[14] Moreira M, Ausas R, Dal Pont S, Pelissari P, Luz A, Pandolfelli V. Towards a single-phase mixed formulation of refractory castables and structural concrete at high temperatures. International Journal of Heat and Mass Transfer. 2021;171:121064.

[15] Kalifa P, Menneteau FD, Quenard D. Spalling and pore pressure in HPC at high temperatures. Cement and Concrete Research. 2000;30(12):1915{1927.

[16] Pimienta P, McNamee RJ, Mindeguia JC. Physical properties and Behaviour of high-Performance Concrete at high Temperature. RILEM State-of-the-Art Reports. 2019.

[17] Dauti D, Tengattini A, Dal Pont S, Toropovs N, Brifaut M, Weber B. Some Observations on Testing Conditions of High-Temperature Experiments on Concrete: An Insight from Neutron Tomography. Transport in Porous Media. 2020:1{12.

[18] Barakat A, Pel L, Adan O. One-dimensional NMR imaging of high temperature first-drying in monolithics. Applied Magnetic Resonance. 2018;49(7):739{753.

[19] Sleiman HC, Tengattini A, Brifaut M, Huet B, Dal Pont S. Simultaneous x-ray and neutron 4D tomographic study of drying-driven hydromechanical behavior of cement-based materials at moderate temperatures. Cement and Concrete Research. 2021;147:106503.

[20] Barakat AJ, Pel L, Krause O, Adan OC. Direct observation of the moisture distribution in calcium aluminate cement and hydratable alumina bonded castables during first-drying: An NMR study. Journal of the American Ceramic Society. 2020;103(3):2101{2113.

[21] Tengattini A, Dal Pont S, Cheikh Sleiman H, Kisuka F, Brifaut M. Quantification of evolving moisture pro les in concrete samples subjected to temperature gradient by means of rapid neutron tomography: Influence of boundary conditions, hygro-thermal loading history and spalling mitigation additives. Strain. 2020;56(6):e12371.

[22] Pileggi R, Pandolfelli V, Paiva A, Gallo J. Novel rheometer for refractory castables. American Ceramic Society Bulletin. 2000;79(1):54{58.

[23] Barsoum MW. Fundamentals of ceramics. New York: McGraw Hill; 1997.

[24] Hidnert P. Thermal expansion of titanium. J Res Natl Bur Stand. 1943;30(1934):101.

[25] Tengattini A, Lenoir N, And o E, Giroud B, Atkins D, Beaucour J, et al. NeXT-Grenoble, the Neutron and X-ray tomograph in Grenoble. Nuclear Instruments and Methods in Physics Research Section A: Accelerators, Spectrometers, Detectors and Associated Equipment. 2020;968:163939.

[26] Savitzky A, Golay MJ. Smoothing and differentiation of data by simplified least squares procedures. Analytical Chemistry. 1964;36(8):1627{1639.

[27] Guerrieri M, Fragomeni S. Spalling of Large-Scale Walls Exposed to a Hydrocarbon Fire. Journal of Materials in Civil Engineering. 2019;31(11):04019249.